# Short-range magnetic interactions and spin-glass behavior in the quasi-2D nickelate $Pr_4Ni_3O_8$

Shangxiong Huangfu[*], Zurab Guguchia[†], Denis Cheptiakov[‡], Xiaofu Zhang[*], Hubertus Luetkens[†], Dariusz Jakub Gawryluk[§], Tian Shang[*], Fabian O. von Rohr[**], Andreas Schilling[*]

[*]Department of Physics, University of Zurich, Winterthurerstrasse 190, CH-8057, Zurich, Switzerland

[†]Laboratory for Muon Spin Spectroscopy (LMU), Paul Scherrer Institute (PSI), Forschungsstrasse 111, CH-5232 Villigen, Switzerland

[‡]Laboratory for Neutron Scattering and Imaging (LNS), Paul Scherrer Institute (PSI), Forschungsstrasse 111, CH-5232 Villigen, Switzerland

[§]Laboratory for Multiscale Materials Experiments (LMX), Paul Scherrer Institute (PSI), Forschungsstrasse 111, CH-5232 Villigen, Switzerland

[**]Department of Chemistry, University of Zurich, Winterthurerstrasse 190, CH-8057, Zurich, Switzerland

*The nickelate $Pr_4Ni_3O_8$ features quasi-two-dimensional layers consisting of three stacked square-planar $NiO_2$ planes, in a similar way to the well-known cuprate superconductors. The mixed-valent nature of Ni and its metallic properties makes it a candidate for potentially unconventional superconductivity. We have synthesized $Pr_4Ni_3O_8$ by topotactic reduction of $Pr_4Ni_3O_{10}$ in 10 % hydrogen gas, and report on measurements of powder-neutron diffraction, magnetization and muon-spin rotation (μSR). We find that $Pr_4Ni_3O_8$ shows complicated spin-glass behavior with a distinct magnetic memory effect in the temperature range from 2 to 300 K and a freezing temperature $T_s \approx 68$ K. Moreover, the analysis of μSR spectra indicates two magnetic processes characterized by remarkably different relaxation rates: a slowly-relaxing signal, resulting from paramagnetic fluctuations of Pr/Ni ions, and a fast-relaxing signal, whose relaxation rate increases substantially below ≈ 70 K which can be ascribed to the presence of short-range correlated regions. We conclude that the complex spin-freezing process in $Pr_4Ni_3O_8$ is governed by these multiple magnetic interactions. It is possible that the complex magnetism in $Pr_4Ni_3O_8$ is detrimental to the occurrence of superconductivity.*

Copper-based high-temperature superconductors show quasi-two-dimensional (2D) structural and electrical properties. They contain $CuO_2$ planes with Cu in a mixed-valent $3d^{9\pm\delta}$ configuration, and exhibit considerable antiferromagnetic correlations [1,2]. Even before [3] but also soon after their discovery [4], structurally similar layered nickelates have been proposed to be potential superconductors [5]. The rare-earth ($Ln$) nickelates $Ln_{n+1}Ni_nO_{2n+2}$ exhibit such a 2D layered structure. They contain $n$ stacked square-planar $NiO_2$ planes, alternating with $Ln$/$O_2$/$Ln$ fluorite-type layers, which is reminiscent to the configuration of the $CuO_2$ planes in electron-doped cuprate superconductors [6-8]. The mixed-valence state of Ni ion between $3d^8$ and $3d^9$ in these nickelates, with a formal Ni valence of $1+1/n$, resembles the situation in heavily hole-doped cuprates [9-12]. Therefore, these nickel oxides have been considered as promising candidates for new superconductors [13]. Indeed, superconductivity has very recently been found in thin films of the "infinite layer" compound $Nd_{1-x}Sr_xNiO_2$ ($n = \infty$) [14-16]. To synthesize these structures, one of the common synthesis techniques is to first prepare $T$-type materials containing $NiO_6$ octahedra, and to subsequently reduce them via a low-temperature topotactic reaction to obtain $T'$-type structures with square-planar $NiO_2$ planes [7,17,18].

The related nickelate $LaNiO_2$ [7,17] is paramagnetic with apparently two different magnetic regimes (above and below $T \approx 150$ K, respectively), with no signs of magnetic order [7]. Likewise, the isostructural $NdNiO_2$ does not show any signs of long-range magnetic order down to $T \approx 1.7$ K [19]. $La_3Ni_2O_6$ is the only known member of the $Ln_{n+1}Ni_nO_{2n+2}$ series with a bilayer ($n = 2$) structure [10]. Nuclear magnetic resonance (NMR) measurements down to 5 K showed the presence of antiferromagnetic spin fluctuations [20], although no long-range magnetic order develops. The trilayer ($n = 3$) versions, $Ln_4Ni_3O_8$ ($Ln$ = La, Pr and Nd/Sm) to be discussed here, have an average Ni valence of $+4/3$. $La_4Ni_3O_8$ shows a sharp phase transition from a high-temperature paramagnetic semiconducting phase to a low-temperature antiferromagnetic-like insulating phase around $T \approx 105$ K [11], which has also been studied by $\mu$SR techniques [21]. NMR measurements revealed 2D antiferromagnetic fluctuations [22], and both magnetic and charge-ordering phenomena were inferred from X-ray and neutron-diffraction (ND) measurements [23,24]. Semiconducting behavior has also been reported for $Nd_{3.5}Sm_{0.5}Ni_3O_8$, with a trend to increasing metallicity under pressure [25].

By very contrast, $Pr_4Ni_3O_8$ is metallic (but not superconducting) at ambient pressure between 2 and 300 K, without any signs for a possible phase transition, neither from transport nor from heat-capacity measurements [12]. At present, very little is known about the magnetic behavior of this compound. We have, therefore, performed a series of powder ND, magnetization and $\mu$SR measurements on $Pr_4Ni_3O_8$ powders.

We firstly synthesized high-quality powdered samples of $Pr_4Ni_3O_{10}$ according to the method described in Refs. [26] and [27]. Subsequent thermo-gravimetric experiments suggest a resulting oxygen content of 7.98(4). The high quality of the thus obtained $Pr_4Ni_3O_8$ powder was confirmed by X-ray diffraction (XRD) and powder-ND measurements (Figs. S1 to S3 in [28-31]) without any indication for impurities within the detection limit. The ND measurements were performed at $T = 1.5$ and 300 K, and the zero-field (ZF) $\mu$SR experiments between 5 and 300 K. The quasi-static (DC) and the alternating current (AC) magnetization were studied from 2 to 350 K and from 2 to 300 K, respectively. To cross-check the magnetic data, some of the $Pr_4Ni_3O_8$ powder was reversibly transformed back into $Pr_4Ni_3O_{10}$. Details about sample preparation and the subsequent characterization are given in the Supplemental Material [28].

ND techniques not only probe purity and quality of crystalline compounds, but can also reveal their magnetic structure. Apart from a shift of the nuclear peaks with temperature, our corresponding refinements [28] do not indicate any long-range ferromagnetic or antiferromagnetic orders in the temperature range between 1.5 and 300 K.

To study the magnetic properties further, we performed DC magnetization measurements by collecting temperature-dependent field-cooling (FC) and zero-field cooling (ZFC) magnetization $M(T)$ data in various external magnetic fields, see Figs. 1(a) and S6 in [28], and by measuring magnetization $M(H)$ loops at different temperatures [Fig. 1(b)]. The magnetization $M$ of $Pr_4Ni_3O_8$ shows a similarly complex behavior as other $T'$-type nickelates [21-23]. While in magnetic fields larger than $\mu_0 H \approx 0.5$ T, $M(T)$ appears to be reversible within the resolution of the measurement, FC and ZFC data gradually separate with decreasing $H$. In $\mu_0 H \approx 0.01$ T, the corresponding $M(T)$ curves are separated up to the highest investigated temperature $T = 350$ K. In $\mu_0 H > 1$ T, a broad, peak-like structure develops that becomes more pronounced with increasing field, in a similar fashion as it has been observed in $La_4Ni_3O_8$ and attributed to the presence of superparamagnetic Ni-particles [10], although the feature is much broader in our $Pr_4Ni_3O_8$ data than in $La_4Ni_3O_8$. The corresponding maximum in $M(T)$ shifts from $T \approx 67$ K for $\mu_0 H = 1$ T to $\approx 85$ K in 7 T. In Figs. 1(b), we show corresponding magnetization $M(H)$ loops taken at different temperatures. The data are hysteretic at all $T$, with coercitives ranging from $\approx 0.044$ T at 2 K to $\approx 0.011$ T at room temperature, and a remanent magnetization of the order of $\approx 1200$ emu $mol^{-1}$ at 2 K and $\approx 550$ emu $mol^{-1}$ at room temperature [lower inset of Fig. 1(b)], respectively, which is reminiscent of ferromagnetic-like behavior.

We can interpret these data as a superposition of a ferromagnetic-like contribution that is saturating in high enough fields, and a paramagnetic term that is linear in $H$ as it has been done, e.g. for $NdNiO_{2+x}$ [19]. In Fig. 2 we show the temperature dependence of both contributions. The high-temperature part of the linear paramagnetic susceptibility above $\approx 160$ K can be well fitted by a Néel-type law, $\chi_{para}(T) = C/(T + \Theta) + \chi_0$, with a Curie-constant $C \approx 6.8$ emu K $Oe^{-1}$ $mol^{-1}$, a Pauli-paramagnetic susceptibility $\chi_0 \approx 10^{-3}$ emu $Oe^{-1}$ $mol^{-1}$, and $\Theta \approx 160$ K. If we assume the free-ion value for the magnetic moment of $Pr^{3+}$ (3.58 $\mu_B$) as it has been found in $Pr_4Ni_3O_{10}$ [26,27], we obtain a residual magnetic moment of $\approx 1.70$ $\mu_B$, which would be consistent with the localized moment of one spin-½ $Ni^+$ ion. By simple valence counting, one would expect the presence of two $Ni^+$ and one $Ni^{2+}$, however, indicating that some $3d$ electrons are delocalized and may lead to the observed metallic behavior. In the high temperature range above $\approx 150$ K, the saturated magnetic moment $M_{sat}$ [inset of Fig. 2] can be well fitted by a $1 - (T/T^*)^{3/2}$ law, with an extrapolated zero-temperature magnetic moment of $\approx 0.22$ $\mu_B$ $mol^{-1}$ and a $T^* \approx 1100$ K.

If this large saturated magnetic moment were due to impurities (e.g., Ni particles as suggested in Ref. [10], with $M_{sat} \approx 0.66$ $\mu_B$ $mol^{-1}$), $\approx 10\%$ of $Pr_4Ni_3O_{10}$ must have been decomposed during the topotactic reduction to $Pr_4Ni_3O_8$ since the original $Pr_4Ni_3O_{10}$ powder showed neither magnetic hysteresis nor saturation. Such a significant decomposition should have manifested itself in our X-ray and ND data, however, which is not the case. To further cross-check our $M(H)$ data, we transformed $Pr_4Ni_3O_8$ back into pure-phase $Pr_4Ni_3O_{10}$, see upper inset of Fig. 1(b) and Supplemental Material [28]. The resulting magnetization $M(H)$ curves are identical to those of the original $Pr_4Ni_3O_{10}$ powder. The ferromagnetic-like hysteretic behavior completely vanished, and it must therefore be intrinsic to $Pr_4Ni_3O_8$.

As the DC magnetization data are clearly hysteretic, we performed additional time-dependent magnetization experiments, namely AC magnetic susceptibility measurements in zero static external magnetic field, and aging and memory experiments for the DC magnetization. We observe a broad peak in the temperature dependent real part of the AC susceptibility $\chi'(T)$ for all AC frequencies $f$ between 2 and 300 Hz, as shown in Fig. 3(a) and S7 [28]. Moreover, the peak positions shift from $\approx 74$ to $\approx 89$ K with the increasing $f$, while the peak amplitudes decrease gradually, indicating a possible spin-glass freezing below the respective peak temperatures. Aging and memory measurements taken in a FC protocol also show a clear time and history dependence of the DC magnetization (see Supplemental Material [28]), which is regarded as an experimental signature for spin-glass behavior [32]. Fig. 3(b) shows the intermittent-stop cooling (ISC) data taken in $\mu_0 H = 5$ mT. In these measurements, $H$ was switched from this value to zero, followed by a waiting period of 5000 s at constant temperature. During this time, the magnetization changed slowly with a time constant of several minutes [upper inset of Fig. 3(b)]. After that process, the magnetic field was switched back to 5 mT and the sample was further cooled down to the next temperature. In spin-glasses, the spin dynamics is known to slow down as the freezing temperature $T_s$ is approached, which impedes the magnetization recovery and leads to large magnetizations steps in such ISC curves [33]. Indeed, these steps are largest between 30 and 50 K, and become smaller with increasing temperature but remain finite up to room temperature [lower inset of Fig. 3(b)]. Although $T_s$ appears to have a finite value around $\approx 70$ K (see below), the continuous slow freezing process persists down to the lowest investigated temperature (5 K). Upon warming the sample continuously up (continuous warming, CW) in the same constant magnetic field after such an ISC process, $M(T)$ exhibits pronounced kink-like features at each of the previous intermittent-stopping temperatures, indicating a distinct memory effect that is characteristic for many spin-glasses, including some nickel oxides [32, 34, 35].

To analyze the frequency dependence of the peaks in $\chi'(T)$ [inset of Fig. 3(a)], we may use a critical dynamical scaling model which predicts a relation between the measuring frequency $f = 2\pi/\tau$ and the peak temperature $T$ in $\chi'(T)$ via $\tau = \tau_0(\frac{T-T_s}{T_s})^{-zv}$, where $T_s$ is the freezing temperature in the limit $\omega \rightarrow 0$ that is determined by the system interactions, $z$ is the dynamic critical exponent, $v$ the exponent for the correlation length, and $\tau_0$ the flip time between relaxation attempts. The solid line in the inset of Fig. 3(a) suggests that the behavior of Pr$_4$Ni$_3$O$_8$ can be well described by this model with $T_s = 68.3 \pm 0.3$ K, $zv = 3.8 \pm 0.2$, and $\tau_0 \approx 10^{-6}$ s. The exponent $zv$ is just in the range of normal glassy systems where $zv = 4$ to $12$ [37]. The flip time $\tau_0$ falls in the order of magnitude $10^{-6}$ s to $10^{-9}$ determined by the atomic spin-flip time [38]. Another way to analyze the frequency dependence of the peaks in $\chi'(T)$ is to apply the Vogel-Fulcher model, which characterizes a continuous freezing crossover into a low-temperature glassy regime [38,39], $\tau = \tau_0 \exp[E_a/k_B(T - T_0)]$, where $E_a$ is an energy barrier, k$_B$ the Boltzmann constant, and $T_0$ a characteristic temperature which is usually below the freezing temperature. The dashed line in the inset of Fig. 3(a) indicates that the same data can be well described by this model as well, with $T_0 = 60.1 \pm 0.8$ K $< T_s$, $E_a/k_B \approx 130$ K and $\tau_0 \approx 10^{-6}$ s. The energy barrier fulfills $E_a/k_B \approx 2T_s$, as observed in other spin glasses [40]. We conclude that Pr$_4$Ni$_3$O$_8$ indeed shows spin-glass behavior with a freezing temperature $T_s \approx 68$ K, which is consistent with the pronounced aging and memory effects that become most significant below $\approx 80$ K

(lower inset of Fig. 3(a)). Below $T_s$, the saturated magnetic moment [inset of Fig. 2] also strongly increases with decreasing temperature down to the lowest investigated temperature.

To further characterize the complex magnetic behavior in Pr$_4$Ni$_3$O$_8$, we performed additional ZF-$\mu$SR measurements, where positive muons are implanted into a sample and serve as an extremely sensitive local probe to detect small internal magnetic fields and ordered magnetic volume fractions in the bulk of magnetic materials. Thus, ZF-$\mu$SR is particularly powerful to study inhomogeneous magnetism in materials. Fig. 4(a) shows corresponding spectra collected between 5 and 300 K. Different from La$_4$Ni$_3$O$_8$ [21], the ZF-$\mu$SR data in Pr$_4$Ni$_3$O$_8$ do not show any oscillations over the whole measured temperature range. This indicates the absence of long-range magnetic order in Pr$_4$Ni$_3$O$_8$ down to 5 K, in agreement with our ND result. The spectra show an initial fast decay and then a much slower one. The best fit to these spectra is achieved with two exponential paramagnetic responses with different relaxation rates [41- 43], $A(t) = A_0 + A_{fast} \exp(-\lambda_{fast} t) + A_{slow} \exp(-\lambda_{slow} t)$. The last two terms represent the fast and the slow relaxation, respectively, with the relaxation rates $\lambda_{fast}$ and $\lambda_{slow}$ for each component, and an additional constant offset $A_0$. In Figures. 4(b) and (c), three different regimes can be clearly distinguished: i) the temperature range 300 - 150 K, in which the relaxation rate of the fast component $\lambda_{fast}$ shows a weak temperature dependence; ii) below $\approx$ 150 K, where $\lambda_{fast}$ starts to increase from $\approx$ 0.8 $\mu s^{-1}$ at $\approx$ 150 K to $\approx$ 3 $\mu s^{-1}$ at $\approx$ 70 K; and iii) below $\approx$ 70 K, where $\lambda_{fast}$ shows a strong increase and reaches an almost saturated value of $\approx$ 38 $\mu s^{-1}$ at 5 K. The rate $\lambda_{slow}$ also shows a weak $T$-dependence down to $\approx$ 150 K, below which it increases from $\approx$ 0.1 to $\approx$ 0.86 $\mu s^{-1}$ at 5 K. We note that the lowest value $\lambda_{fast}$ is still much higher than the corresponding value of $\lambda_{slow}$. Such a remarkable difference rules out the origin of these two decays being due to slight inhomogeneities, such as a compositional distribution within the polycrystalline sample, temperature gradients, or any other small spatial variations. While it is plausible to identify $\lambda_{slow}$ as originating from the paramagnetism in Pr$_4$Ni$_3$O$_8$, i.e., the fluctuations of Pr/Ni ions in the paramagnetic state, the strong quantitative difference between the relaxation rates $\lambda_{\text{fast}}$ and $\lambda_{\text{slow}}$ indicates the presence of another source of fast muon depolarization, which is different from pure paramagnetism. In general, a fast-relaxing decay indicates the presence of strong spin-spin correlations. Stronger magnetic correlations between the spins slow down the spin fluctuations, leading to a significant rise of the relaxation rate. Thus, our $\mu$SR results indicate the slowing down of spin fluctuations starting already below $\approx$ 150 K, and the formation of strong short-range order correlations below $\approx$ 70 K. This is consistent with the formation of a spin-glass-like state below this temperature, most likely due to the presence of short-range ordered magnetic clusters or islands, and we estimate the volume fraction of this state to $\approx$ 40 % at 5 K.

The quantities $\chi_{\text{para}}$, $\lambda_{\text{fast}}$ and $\lambda_{\text{slow}}$ all show changes in their $T$-dependences already at $\approx$ 150 K besides those at $T_s$. This may indicate a further transition (or crossover) taking place around this temperature, the origin of which is unknown to us at present. The change of the Néel-type magnetic susceptibility $\chi_{para}(T)$ below $T \approx \Theta \approx 160$ K is too large to be solely ascribed to the nickel subsystem. It is conceivable that a that large volume fraction in the system $Pr_4Ni_3O_8$ behaves as a strongly correlated paramagnet, in which $Pr^{3+}$ has a very strong tendency to form short-range-ordered magnetic clusters, although no long-range magnetic order develops.

We cannot clearly distinguish between contributions from Pr and Ni ions in our $\mu$SR experiments, and it is likely that both contribute to the $\mu$SR relaxation rates. Therefore, we cannot draw further conclusions about the microscopic details of the spin-glass formation. The lack of magnetic order may be due to a certain magnetic frustration, eventually leading to the formation of a spin glass of either Ni or Pr magnetic moments, or both. However, the relatively small difference between FC and ZFC magnetization data (with $M_{\text{sat}}$ as an upper limit, see Figs. 1 and 2) rather points to a glassy state within the Ni subsystem.

To summarize, the powder-ND and $\mu$SR measurements show no evidence for long-range magnetic order in $Pr_4Ni_3O_8$. Our magnetization data clearly demonstrate a spin-glass behavior with a freezing temperature $T_s \approx$ 68 K and a distinct magnetic memory effect at all temperatures. Two magnetic processes are characterized by remarkably different relaxation rates: a slowly-relaxing signal, resulting from paramagnetic fluctuations present at all temperatures, and a rapidly growing fast-relaxing signal due to the presence of short-range correlated regions in the glassy state below $T_s$. This complicated magnetic behavior may prohibit the occurrence of superconductivity in $Pr_4Ni_3O_8$. However, it is conceivable that chemical doping (i.e., a further change of the Ni valence state) suppresses the spin-glass state and eventually renders the trilayer $T'$-type nickelate $Pr_4Ni_3O_8$ superconducting [13].

**Acknowledgements**
This work was supported by the Swiss National Science Foundation under Grants No.20-175554, 206021-163997 and PZ00P2-174015. Neutron diffraction measurements have been supported by the European Commission under the 7th Framework Programme through the 'Research Infrastructures' action of the 'Capacities' Programme, NMI3-II Grant No.283883.

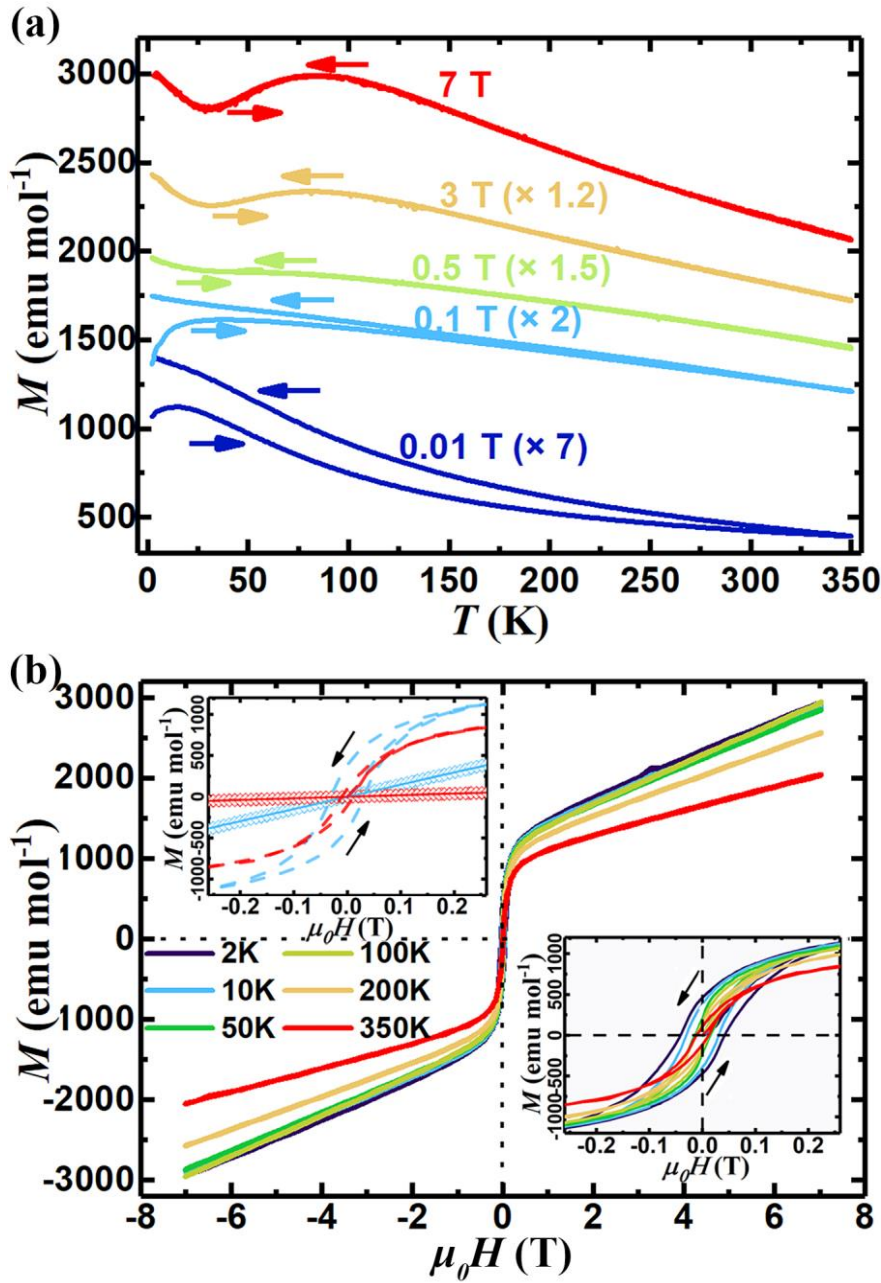

Fig.1 (a) Temperature-dependent ZFC and FC magnetization $M(T)$ measurements on $Pr_4Ni_3O_8$ in various external magnetic fields up to $\mu_0 H = 7$ T; arrows indicate the direction of the measurements; some data have been multiplied by a factor for clarity; (b) Isotherm magnetization $M(H)$ loops with magnetic fields ranging from -7 to 7 T at temperatures between 2 and 350 K ; the lower inset is in an expanded scale to show the hysteretic behavior in low magnetic fields, while the upper inset shows corresponding data for 10 and 350 K for $Pr_4Ni_3O_8$ (dashed lines) in comparison with back-oxygenated $Pr_4Ni_3O_{10}$ from $Pr_4Ni_3O_8$ (open symbols with no detectable hysteresis) and the original $Pr_4Ni_3O_{10}$ powder (solid lines); arrows indicate the direction of the measurements.

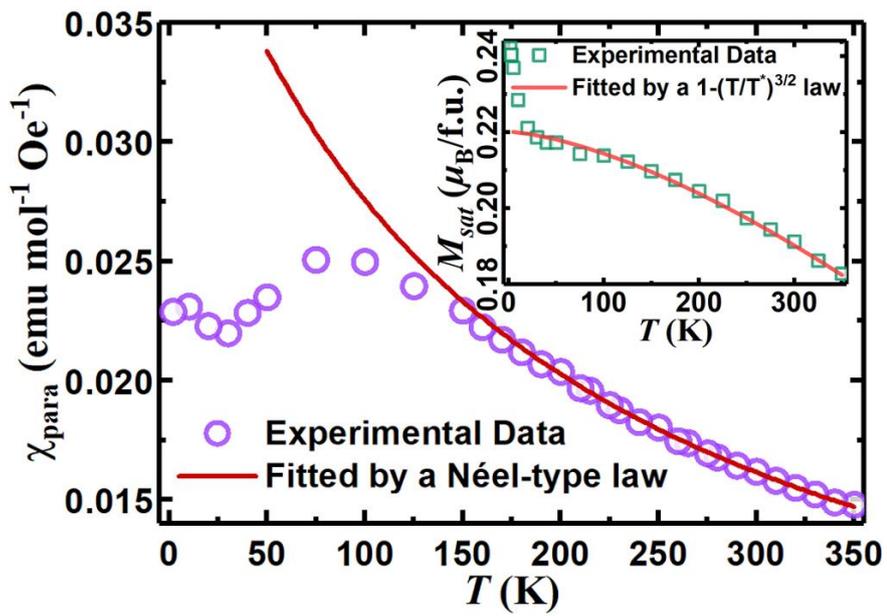

Fig.2 Paramagnetic susceptibility as calculated from the high-field linear part of $M(H)$ and fitted to a Néel-type law; the inset shows the high-field saturated magnetic moment obtained by subtracting the paramagnetic magnetization from $M(H)$.

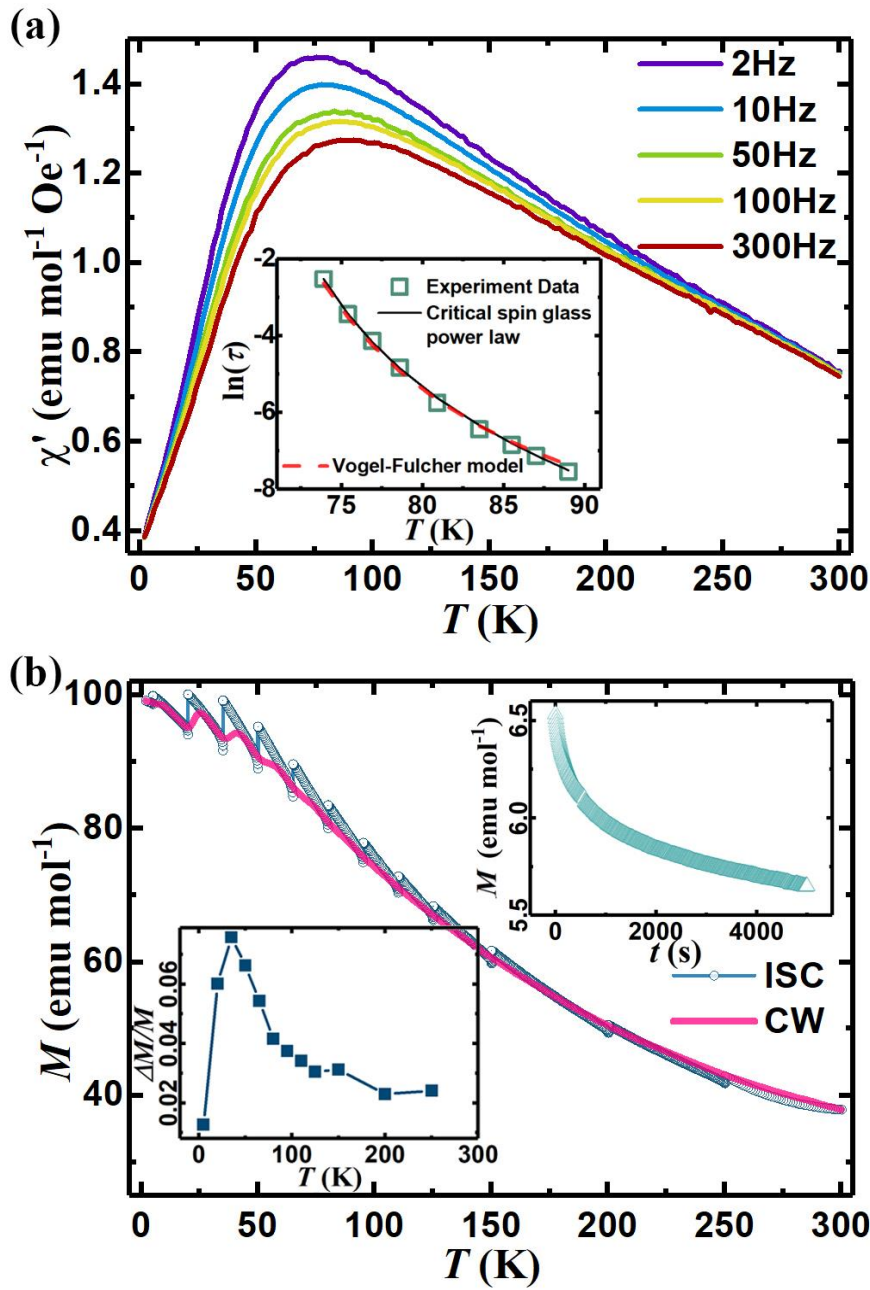

Fig.3 (a) Real part ($\chi'$) of the AC susceptibility as a function of temperature and frequency from 2 to 300 Hz; the inset of (a) shows the measurement period $\tau = 1/2\pi f$ (where $f$ is the frequency of applied AC magnetic field) *vs.* peak temperatures $T$ of the real part of the AC magnetic susceptibility $\chi'$; (b) Aging and memory measurements according to intermittent-stop cooling (ISC) and continuous warming-up (CW) processes, respectively, in an external field of 5 mT, with a waiting time in the ISC process of 5000 s; the upper inset of (b) shows the time dependence of the magnetic moment at 35 K after switching off the magnetic field; the lower inset of (b) shows the total change of the magnetic moment during the intermittent-stop.

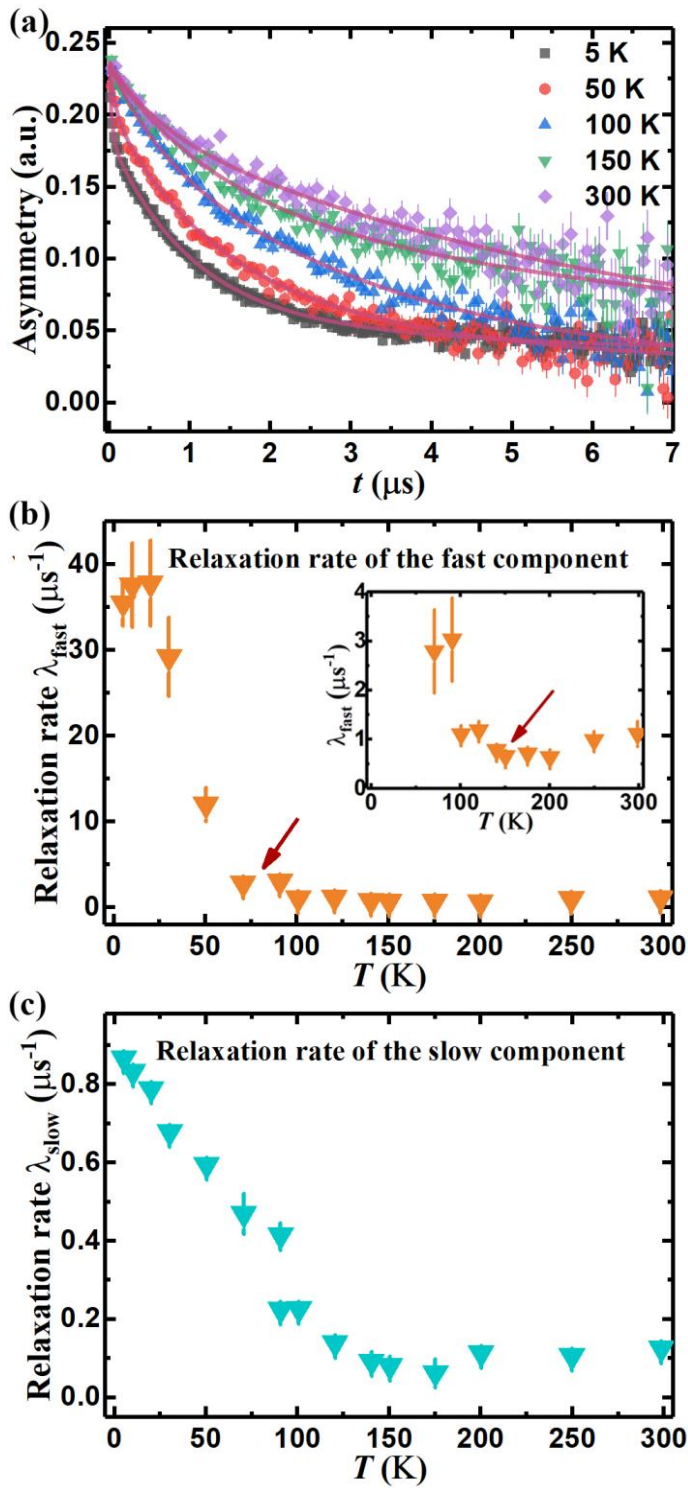

Fig. 4 (a) Zero-field μSR spectra at 5, 50, 100, 150, and 300K; the relaxation rates of the fast component (b) and of the slow component (c) as functions of temperature are calculated from the corresponding μSR asymmetry spectra; the inset of (b) is in an expanded scale to show the fast component relaxation rate at high temperatures; arrows mark the change of the relaxation rate.

# Short-range magnetic interactions and spin-glass behavior in the quasi-2D nickelate $Pr_4Ni_3O_8$

Shangxiong Huangfu*, Zurab Guguchia†, Denis Cheptiakov‡, Xiaofu Zhang*, Hubertus Luetkens†, Dariusz Jakub Gawryluk§, Tian Shang*, Fabian O. von Rohr**, Andreas Schilling*

*Department of Physics, University of Zurich, Winterthurerstrasse 190 CH-8057 Zurich Switzerland

†Laboratory for Muon Spin Spectroscopy (LMU), Paul Scherrer Institute (PSI), Forschungsstrasse 111, CH-5232 Villigen, Switzerland

‡Laboratory for Neutron Scattering and Imaging (LNS), Paul Scherrer Institute (PSI), Forschungsstrasse 111, CH-5232 Villigen, Switzerland

§Laboratory for Multiscale Materials Experiments (LMX), Paul Scherrer Institute (PSI), Forschungsstrasse 111, CH-5232 Villigen, Switzerland

**Department of Chemistry, University of Zurich, Winterthurerstrasse 190 CH-8057 Zurich Switzerland

***Synthesis***

Powdered samples of $Pr_4Ni_3O_{10}$ were synthesized by a citric acid assisted sol-gel method. The reactants $Pr_6O_{11}$ (99.9%; *Sigma-Aldrich*) and NiO (99.99%; *Sigma-Aldrich*) were weighted in stoichiometric ratios. Then, nitric acid (65% for analysis; *Emsure*) was used for dissolving these oxides, and a clear green liquid was obtained. After adding citric acid monohydrate ($C_6H_8O_7*H_2O$; 99.5%; *Emsure*) in molar ratio 1:1 with respect to the molar of cations, the resulting liquid was heated at around 300 ˚C on a heating plate, dried and decomposed into a dark brown powder. Finally, an ultrafine and homogeneous powder of $Pr_4Ni_3O_{10}$ was obtained by annealing the precursor powder at 1100 ˚C in flowing oxygen for 24 hours.

In the topotactic reduction, 10 mol % of $H_2$ mixed with 90 mol % of $N_2$ was used to reduce $Pr_4Ni_3O_{10}$ to $Pr_4Ni_3O_8$ at 360 ºC for 18 hours.

The oxygen content of the obtained sample is 7.98(4) as determined by thermogravimetric reduction with 10% of $H_2/N_2$ gas.

To cross-check the magnetic data, some of the obtained $Pr_4Ni_3O_8$ powder was annealed at 600 ˚C in air for 18 hours. Powder XRD shows that the $Pr_4Ni_3O_8$ transformed back to phase-pure $Pr_4Ni_3O_{10}$ (see Fig.S5).

*Powder X-ray and neutron diffraction*

Powder X-Ray diffraction (PXRD) data were collected at room temperature in transmission mode using a Stoe Stadi P diffractometer and Cu$K_{\alpha 1}$ radiation (Ge(111) monochromator) with a DECTRIS MYTHEN 1K detector. Powder neutron-diffraction measurements were done at the High-Resolution Powder Diffractometer for Thermal Neutrons (HRPT) at the Swiss spallation neutron source SINQ, Paul Scherrer Institute, Villigen, Switzerland. A neutron wavelength $\lambda$ = 2.45 Å was used for the measurements at temperatures $T$ = 1.5 K and 300K.

All of the reflections were indexed with a tetragonal cell in the space group *I4/mmm* (No 139) for $Pr_4Ni_3O_8$ and with a monoclinic cell in the space group *P2$_1$/a* (No 14) for $Pr_4Ni_3O_{10}$. The Rietveld refinement analysis [29] of the diffraction patterns was performed with the package FULLPROF SUITE [30] (version March-2019). The structural model was taken from the single-crystal X-Ray diffraction refinement. The refined parameters include: scale factor, zero shift, transparency, lattice parameters, Pr and Ni atomic positions (in addition the O positions for the neutron data), and peak shapes as a Thompson–Cox–Hastings pseudo-Voigt function. A preferred orientation correction as a Modified March function was implemented in the analysis. The corresponding results are tabulated in Tables SI-SIV.

*Muon Spin Rotation*

$\mu^+$SR experiments were carried out with the General Purpose Surface-Muon Instrument (GPS) [31] at the piM3.2 muon beam line, Swiss Muon Source, PSI, Switzerland, in the temperature range of 5 K to 300 K without external magnetic field.

*Magnetization measurements*

The magnetic properties of $Pr_4Ni_3O_8$ were studied with a Magnetic Properties Measurement System (MPMS 3, *Quantum Design Inc.*), equipped with an alternating current (AC) magnetic measurement option. The temperature dependent magnetization $M(T)$ was measured between 2 K and 350 K, with external fields $B$ = 0.01 T, 0.1 T, 0.2 T, 0.5 T, 1 T, 3 T, 5 T and 7 T, respectively. The field-dependence of the magnetization $M(H)$ was measured up to 7 T from 2 K to 350K. Moreover, FC memory and aging effects were measured by intermittent-stop cooling and continuous warming memory processes. In the intermittent-stop cooling process, the sample was cooled down from 300 K to 2 K in a field of 0.005 T with intermittent stops at 250 K, 200 K, 150 K, 125 K, 110 K, 95 K, 80 K, 65 K, 50 K, 35 K, 20 K, 5 K respectively. The field was switched off for 5000 s during stops, and then ramped up to 5 mT before cooling resumed. In the continuous warming memory process, the sample was measured in a field of 5 mT continuously from 2 K to 300 K. AC magnetization measurements were done from 2 K to 300K in zero direct magnetic field with peak amplitudes of 0.7 mT and frequencies of 2 Hz, 5 Hz, 20 Hz, 50 Hz, 100 Hz, 150 Hz, 200 Hz, 250 Hz, 300 Hz, respectively.

Table SI Crystallographic parameters obtained by the Rietveld refinements from different diffraction data for $Pr_4Ni_3O_8$

|  | a (Å) | c (Å) | $R_p$ (%) | $R_{wp}$ (%) | $\chi^2$ |
|---|---|---|---|---|---|
| Powder X-ray diffraction at room temperature | 3.9263(1) | 25.487(1) | 1.97 | 3.08 | 2.18 |
| Powder neutron diffraction at 1.5 K | 3.9193(2) | 25.378(1) | 4.33 | 5.69 | 3.99 |
| Powder neutron diffraction at 300 K | 3.9262(2) | 25.484(2) | 4.37 | 5.80 | 2.67 |

Table SII: Atomic positions at 1.5 K from the refinement of powder neutron diffraction for $Pr_4Ni_3O_8$

| Atom | Wyckoff site | x | y | z | occu. |
|---|---|---|---|---|---|
| Pr1 | 4e | 1 | 0 | 0.4354(2) | 1 |
| Pr2 | 4e | 1 | 1 | 0.2994(2) | 1 |
| Ni1 | 4e | 0.5 | 0.5 | 0.3738(2) | 1 |
| Ni2 | 2a | 0.5 | 0.5 | 0.5 | 1 |
| O1 | 4c | 0.5 | 0 | 0.5 | 1 |
| O2 | 8g | 1 | 0.5 | 0.3743(2) | 1 |
| O3 | 4d | 0.5 | 1 | 0.25 | 1 |

Table SIII: Atomic positions at 300 K from the refinement of powder neutron diffraction for $Pr_4Ni_3O_8$

| Atom | Wyckoff site | x | y | z | occu. |
|---|---|---|---|---|---|
| Pr1 | 4e | 1 | 0 | 0.4349(3) | 1 |
| Pr2 | 4e | 1 | 1 | 0.2987(3) | 1 |
| Ni1 | 4e | 0.5 | 0.5 | 0.3749(2) | 1 |
| Ni2 | 2a | 0.5 | 0.5 | 0.5 | 1 |
| O1 | 4c | 0.5 | 0 | 0.5 | 1 |
| O2 | 8g | 1 | 0.5 | 0.3745(2) | 1 |
| O3 | 4d | 0.5 | 1 | 0.25 | 1 |

Table SIV Crystallographic parameters obtained by the Rietveld refinements from XRD for $Pr_4Ni_3O_{10}$

|  | a (Å) | b (Å) | c (Å) | β (°) | $R_p$ (%) | $R_{wp}$ (%) | $\chi^2$ |
|---|---|---|---|---|---|---|---|
| Original $Pr_4Ni_3O_{10}$ | 5.3720(3) | 5.4589(3) | 14.0160(7) | 100.814(1) | 2.33 | 3.59 | 5.28 |
| Back-oxygenated $Pr_4Ni_3O_{10}$ | 5.3780(4) | 5.4670(4) | 14.0329(9) | 100.792(1) | 2.18 | 3.04 | 1.35 |

*Diffraction patterns*

For all diffraction patterns shown below, the black lines corresponds to the best fits from the Rietveld refinement analyses, the vertical marks denote the ideal Bragg peak positions, and the bottom, grey lines represent the difference between experimental and calculated data.

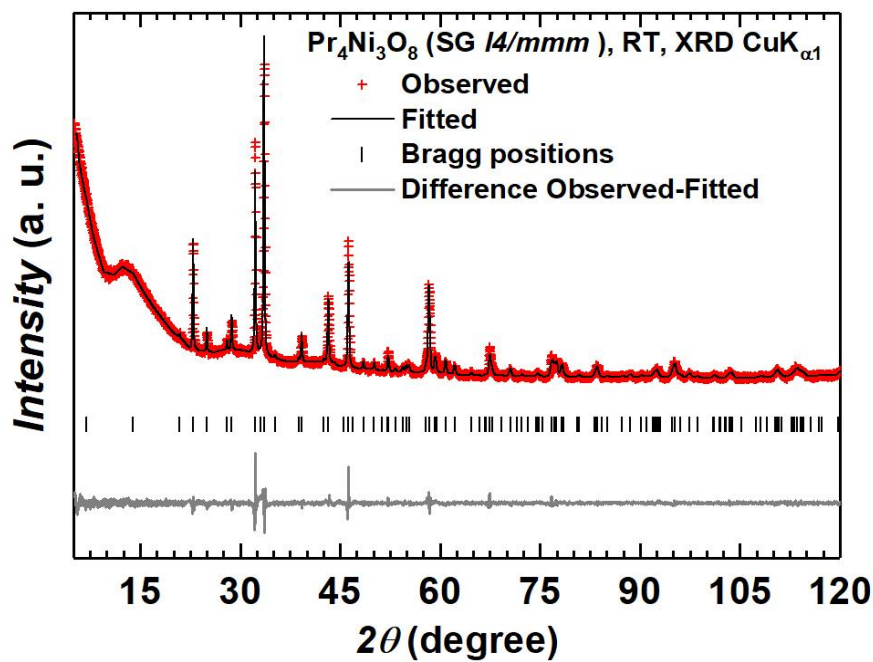

Fig. S1 Experimental powder XRD (Cu$K_{\alpha1}$ radiation) pattern (red crosses) for a powder sample of Pr$_4$Ni$_3$O$_8$ at room temperature.

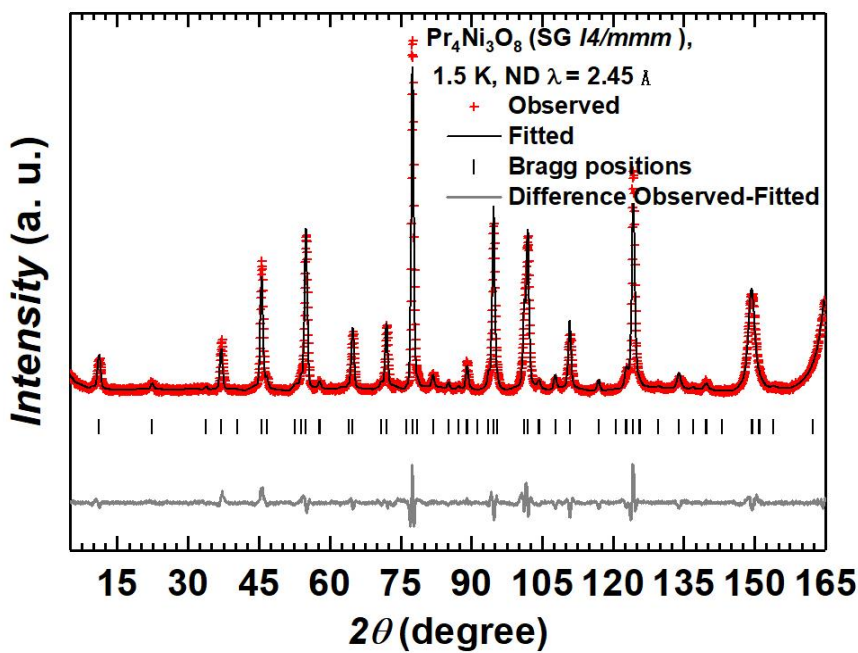

Fig. S2 Experimental powder neutron-diffraction ($\lambda$ = 2.45 Å) pattern (red crosses) for a powder sample of Pr$_4$Ni$_3$O$_8$ at 1.5 K.

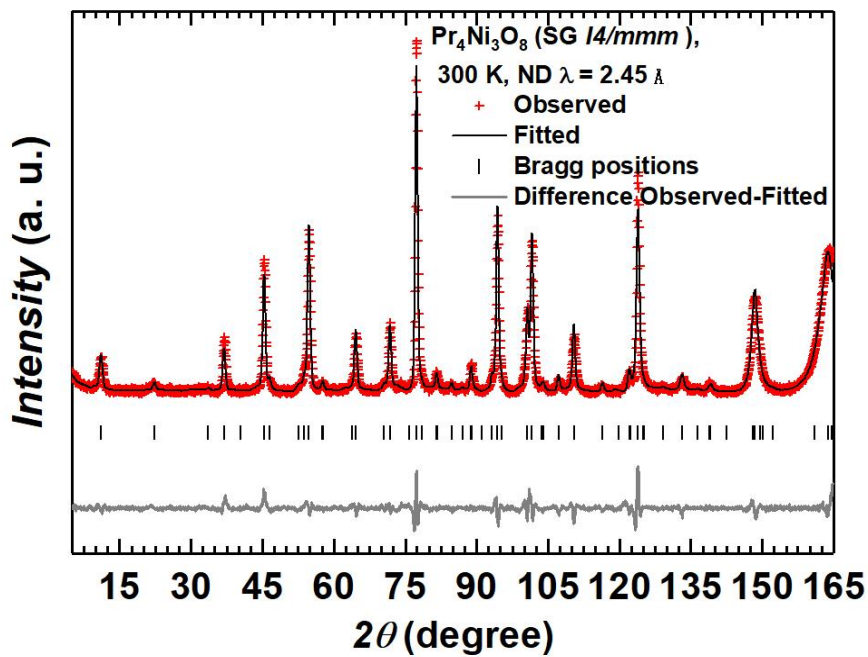

Fig. S3 Experimental powder ND ($\lambda$ = 2.45 Å) pattern (red crosses) for a powder sample of $Pr_4Ni_3O_8$ at 300 K.

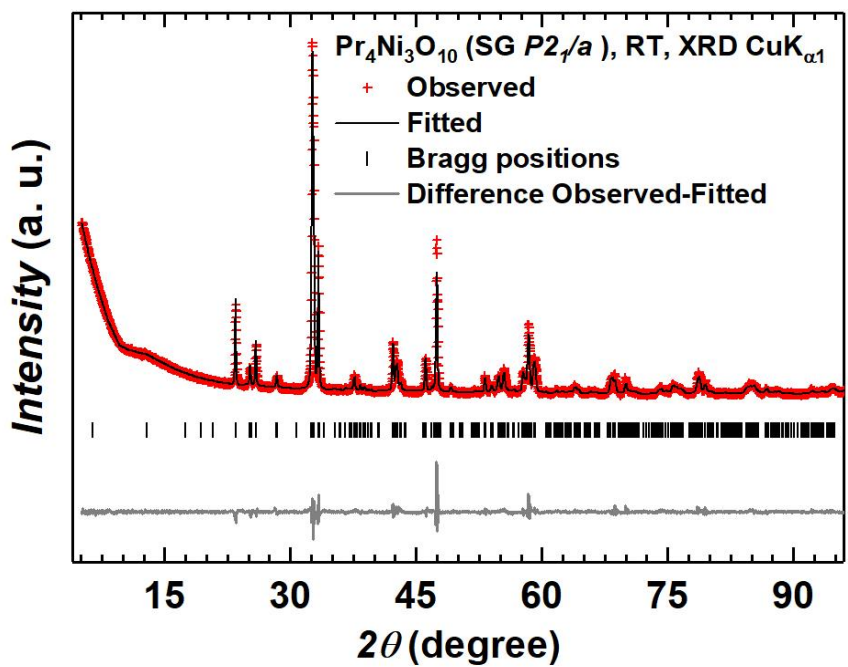

Fig. S4 Experimental powder XRD (CuK$_{\alpha1}$ radiation) pattern (red crosses) of the original $Pr_4Ni_3O_{10}$ powder at room temperature.

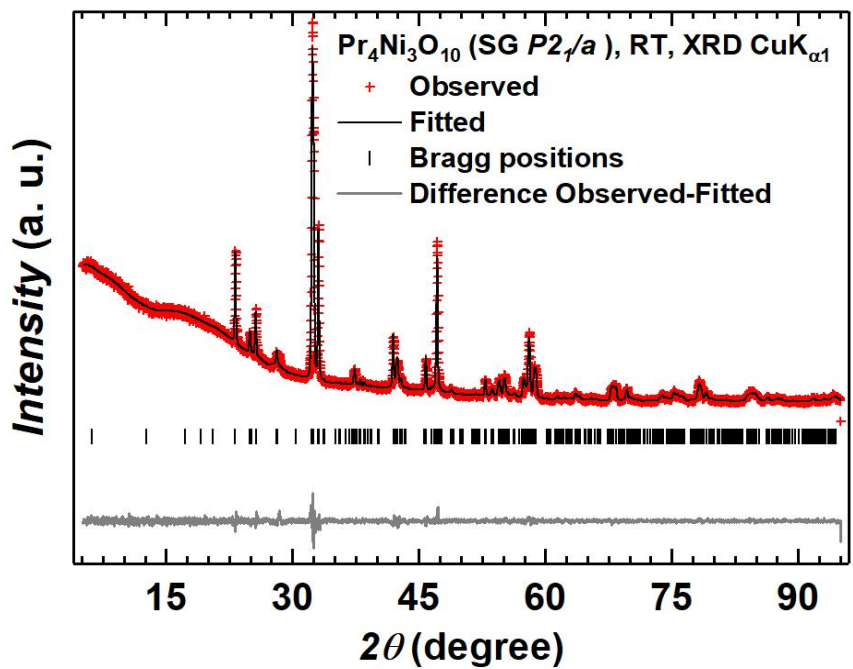

Fig. S5 Experimental powder XRD (CuK$_{\alpha1}$ radiation) pattern (red crosses) of the back-oxygenated Pr$_4$Ni$_3$O$_{10}$ powder at room temperature.

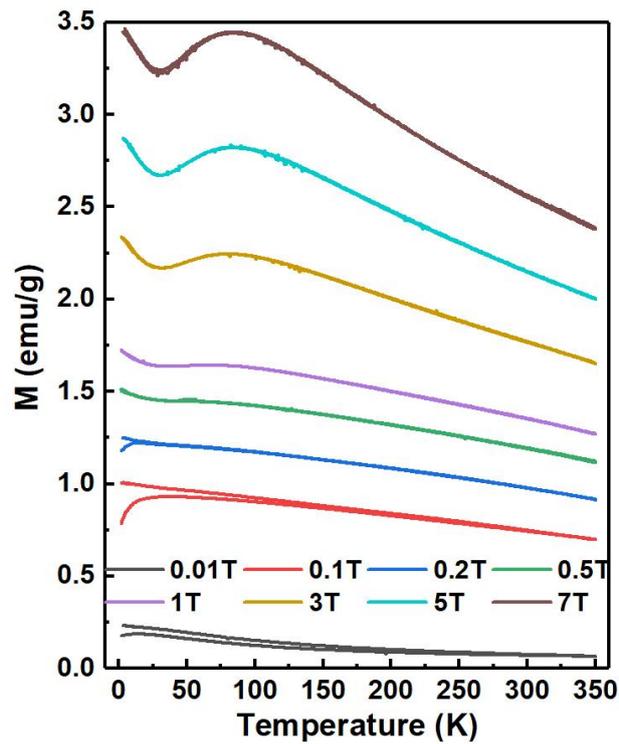

Fig. S6 Temperature dependent ZFC and FC magnetization data in external magnetic fields of 0.01 T, 0.1 T, 0.2 T, 0.5 T, 1 T, 3 T, 5 T and 7 T, respectively.

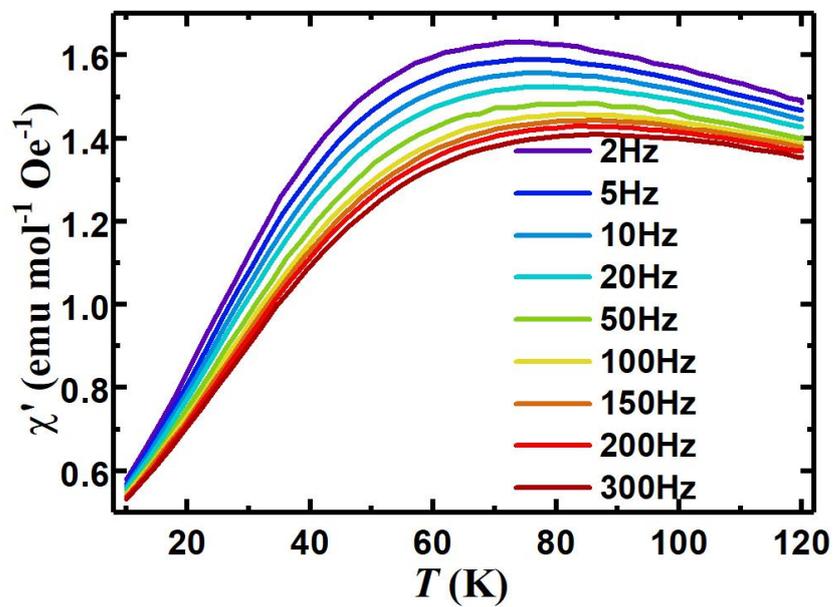

Fig.S7 Real part of the AC magnetic susceptibility ($\chi'$) as a function of temperature with frequencies ranging from 2 Hz to 300 Hz in the temperature range between 10 K and 120 K.